\documentclass[twocolumn,showpacs,amsmath,amssymb,pra]{revtex4}
\usepackage{amsxtra}
\usepackage{amssymb}
\usepackage{amsmath}
\usepackage{graphicx}
\usepackage[bf]{subfigure}

\newcommand {\rhovec}{\ensuremath \boldsymbol{\rho}}
\newcommand {\kvec}{\ensuremath \boldsymbol{k}}

\begin{document}
\title{Signal-to-noise ratio of Gaussian-state ghost imaging}
\date{\today}
\author{Baris I. Erkmen\footnote{Present address:  Jet Propulsion Laboratory, Pasadena, California 91109, USA}}\email{baris.i.erkmen@jpl.nasa.gov}
\author{Jeffrey H. Shapiro}
\affiliation{Massachusetts Institute of Technology, Research Laboratory of Electronics, Cambridge, Massachusetts 02139, USA}

\begin{abstract} 
The signal-to-noise ratios (SNRs) of three Gaussian-state ghost imaging configurations---distinguished by the nature of their light sources---are derived.  Two use classical-state light, specifically a joint signal-reference field state that has either the maximum phase-insensitive or the maximum phase-sensitive cross correlation consistent with having a proper $P$ representation.  The third uses nonclassical light, in particular an entangled signal-reference field state with the maximum phase-sensitive cross correlation permitted by quantum mechanics.  Analytic SNR expressions are developed for the near-field and far-field regimes, within which simple asymptotic approximations are presented for low-brightness and high-brightness sources. A high-brightness thermal-state (classical phase-insensitive state) source will typically achieve a higher SNR than a biphoton-state (low-brightness, low-flux limit of the entangled-state) source, when all other system parameters are equal for the two systems.  With high efficiency photon-number resolving detectors, a low-brightness, high-flux entangled-state source may achieve a higher SNR than that obtained with a high-brightness thermal-state source.  
\end{abstract}
\maketitle 

\section{Introduction}
Ghost imaging is a transverse imaging modality that exploits the cross correlation between two photocurrents, arising from the detection of two distinct but highly-correlated optical beams, to image an object \cite{Gatti:one,Gatti:two,Gatti:three,Cai:one,Cai:two}.  One beam illuminates the object prior to detection by a single-pixel (bucket) detector, while the other undergoes only free-space diffraction before being detected by a high spatial resolution (scanning pinhole or CCD camera) detector.  Ghost imaging was initially demonstrated with biphoton-state light obtained from spontaneous parametric downconversion \cite{Pittman}, which requires a quantum-mechanical description for its photodetection statistics.  Subsequent demonstrations with thermal-state light \cite{Valencia,Ferri}, which admits to a semiclassical interpretation of its photodetection statistics, has generated interest in applying ghost imaging to remote-sensing applications \cite{Meyers:ToySoldier}.

In \cite{ErkmenShapiro:GI} we developed a Gaussian-state framework for the analysis of ghost imaging that provides a unified treatment of biphoton-state and thermal-state illumination~\cite{footnote1}.  There we also introduced a classical-state source of maximum phase-sensitive cross correlation whose ghost-imaging characteristics are most similar to those obtained with biphoton-state illumination. For all of these sources we determined the near-field and far-field image resolution they afford in lensless ghost imaging, and we quantified the low cross-correlation contrast seen with classical Gaussian-state sources and the significant advantage, in this regard, that accrues from use of the biphoton state.  We did not, however, address the signal-to-noise (SNR) behavior of these ghost imagers, although we noted the relevance of having high cross-correlation contrast to achieving high SNR. This relation between SNR and contrast (visibility) has been noted in other earlier treatments of ghost imaging as well, e.g., \cite{Gatti:four}.

 The low cross-correlation contrast of classical-state ghost images---which originates from the appreciable featureless background in which the desired image is embedded---is easily remedied by forming cross-covariance images, rather than cross-correlation images \cite{Gatti:two,Gatti:three,ErkmenShapiro:GI}.  This can be accomplished by {\sc ac}-coupling the photocurrents into a correlator, as was done in \cite{Scarcelli}, or by background subtraction.  Nevertheless, these techniques do not eliminate the noise (shot noise and excess noise) associated with the featureless background, which affects the integration time needed to obtain an accurate cross-covariance estimate.  Therefore it is important to quantify the performance of classical and quantum ghost imagers via their signal-to-noise ratios.  Furthermore, pursuing closed-form analytic expressions for their SNRs is beneficial in identifying the most critical source and detector parameters that impact image quality.  Several valuable contributions have been made toward this end \cite{SalehTeich:GINoise,ChangHan:GISNR}, but the complexity of the variance expression for the image estimate has thus far prevented a rigorous treatment of ghost-image SNR behavior. 

In this paper, we shall utilize our previously-developed Gaussian-state framework to derive tractable analytical expressions for the SNRs of three lensless ghost imagers, whose configurations are distinguished by the nature of their light sources.  Two use classical-state light, specifically a joint signal-reference field state that has either the maximum phase-insensitive or the maximum phase-sensitive cross correlation consistent with having a proper $P$ representation.  The third uses nonclassical light, in particular an entangled signal-reference field state with the maximum phase-sensitive cross correlation permitted by quantum mechanics.   Because the low-flux, low-brightness limit of the last state reduces to vacuum plus a weak biphoton component, our analysis encompasses biphoton-state ghost imagers.   The rest of the paper is organized as follows.  In Sec.~\ref{analysis} we establish our notation and list the general assumptions used in our analysis. Then, for each source, we develop its ghost-image SNR expression and its low-brightness and high-brightness asymptotic behavior in both the near-field and far-field regimes.  In Sec.~\ref{IntTimes} we compare the required averaging times for each source to achieve a desired SNR value.  We conclude, in Sec.~\ref{discussion}, with a summary and discussion of our results.

\section{Analysis}
\label{analysis}

\begin{figure}
\begin{center}
\includegraphics[width= 3in]{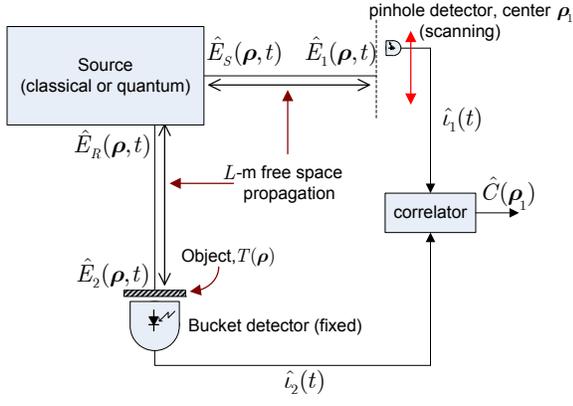}
\end{center}
\caption{(Color online) Lensless ghost imaging configuration.} \label{GI:propagation}
\end{figure}

The lensless ghost imaging configuration that we will consider is shown in Fig.~\ref{GI:propagation}~\cite{footnote2}.    Here, $\hat{E}_{S}(\rhovec,t)e^{-i \omega_{0}t}$ and $\hat{E}_{R}(\rhovec,t)e^{-i \omega_{0}t}$ are scalar, positive-frequency, paraxial, signal ($S$) and reference ($R$) source-field operators, each with center frequency $\omega_{0}$ and $\sqrt{\text{photons}/{\rm m}^2 {\rm s}}$ units.  These operators obey the canonical commutation relations \cite{YuenShapiroPartI} 
\begin{align}
[\hat{E}_{m}(\rhovec_{1},t_{1}),\hat{E}_{\ell}(\rhovec_{2},t_{2})]&=0 
 \label{commutators:1}\\[.05in]
[\hat{E}_{m}(\rhovec_{1},t_{1}),\hat{E}_{\ell}^{\dagger}(\rhovec_{2},t_{2})]&=\delta_{m,\ell}\, \delta(\rhovec_{1} - \rhovec_{2}) \delta(t_{1} - t_{2})\,, \label{commutators:2}
\end{align}
for $m,\ell = S,R$.  
The joint signal-reference source states we shall employ will all be zero-mean jointly-Gaussian states in which the signal and reference beams have identical coherence-separable Gaussian-Schell model phase-insensitive autocorrelation functions given by 
\begin{eqnarray}
\lefteqn{\langle \hat{E}_{x}^{\dagger}(\rhovec_{1}, t_{1}) \hat{E}_{x}(\rhovec_{2}, t_{2}) \rangle  = }
\nonumber \\[.1in]
&& \frac{2 P}{\pi a_{0}^2} e^{-(|\rhovec_{1}|^2+ |\rhovec_{2}|^2)/a_{0}^2 - |\rhovec_{2} - \rhovec_{1}|^2/2 \rho_{0}^2 } e^{-(t_2 - t_1)^2/2T_{0}^{2}} \label{GS:PISauto}
\end{eqnarray}
for $x=S, R$, but vanishing phase-sensitive autocorrelations. Each source will therefore be distinguished by its (phase-insensitive and phase-sensitive) cross-correlation functions, which will be specified in the subsections that follow.  In \eqref{GS:PISauto}:  $P$  denotes the photon flux of the signal and reference; $a_{0}$ is their beam radius; $\rho_{0}$ is their coherence radius, which is assumed to satisfy the low-coherence condition $\rho_{0} \ll a_{0}$; and $T_0$ is their coherence time. 

The source-plane signal and reference field operators both undergo quasimonochromatic, paraxial diffraction over $L$-m-long free-space paths, yielding detection-plane field operators $\hat{E}_{1}(\rhovec,t)$ and $\hat{E}_{2 }(\rhovec,t)$ respectively. These detection-plane field operators are also in a zero-mean jointly-Gaussian state, whose second-order correlation functions can be derived from their source-plane counterparts and the free-space Green's function \cite{ErkmenShapiro:GI,YuenShapiroPartI}.  The first field, $\hat{E}_{1}(\rhovec,t)$, illuminates a quantum-limited pinhole photodetector whose photosensitive region is centered at the transverse coordinate $\rhovec_{1}$, and whose photosensitive area $A_1$ is smaller than the coherence area of the impinging field state. The second field, $\hat{E}_{2}(\rhovec,t)$, illuminates a field-transmission mask $T(\rhovec)$  that is located immediately in front of a quantum-limited bucket photodetector which collects all light transmitted through the mask \cite{footnote3}. 

The photodetectors are assumed to have identical sub-unity quantum efficiency $\eta$ and no dark current or electronic thermal noise. Furthermore, finite-bandwidth postdetection filters {\sc ac}-couple the photocurrents into the correlator block from Fig.~\ref{GI:propagation}, so that a background-free ghost image is obtained. For analytic simplicity, we assume that the composite baseband frequency response of the photodetectors and  their {\sc ac}-coupling is given by the difference of two Gaussian functions \cite{footnote4},
\begin{equation}
H_{B}(\Omega) = \mathcal{F}[ h_{B}(t) ] =  e^{- 2 \Omega^{2}/\Omega_{B}^{2} } - e^{-2\Omega^{2}/ \Omega_{N}}  \,,
\end{equation}
where $\Omega_{B}$ is the baseband bandwidth of the detector (taken at the $e^{-2}$ attenuation level of the frequency response), $\Omega_{N} \ll \Omega_{B}$ is the stopband bandwidth of the {\sc ac}-coupling notch around $\Omega =0$,  and $\mathcal{F}[h_B(t)]$ denotes the Fourier transform of the composite filter's impulse response, $h_B(t)$. In order to minimize suppression of the baseband photocurrent fluctuations---whose cross correlation yields the ghost image---the notch bandwidth will be taken to be much smaller than the bandwidth of the impinging fields, i.e., $\Omega_{N} T_{0} \ll 1$ will be assumed in all that follows. 

The ghost image at the transverse coordinate $\rhovec_1$ is formed by time-averaging the product of the detector photocurrents, which is equivalent to a measurement of the quantum operator
\begin{equation}
\hat{C}(\rhovec_{1}) = \frac{1}{T_{I}} \int_{-T_{I}/2}^{T_{I}/2} dt\, \hat{\imath}_{1}(t) \hat{\imath}_{2}(t)\,,
\end{equation}
where \cite{footnote5}
\begin{equation}
\hat{\imath}_{m}(t) =  q \! \int\! {\mathrm d}\tau \int_{\mathcal{A}_{m}} \! {\rm d}\rhovec \,\hat{E}_{m}^{'\dagger}(\rhovec,\tau) \hat{E}_{m}'(\rhovec,\tau) h_{B}(t-\tau), 
  \end{equation}
for $m=1,2$, with $q$ being the electron charge, $\mathcal{A}_m$ the photosensitive region of detector $m$, and $T_{I}$ the duration of the averaging interval. 
 The field operators appearing in these photocurrent operators are 
 \begin{eqnarray}
 \lefteqn{\hat{E}_m'(\rhovec,t) \equiv } \nonumber \\[.1in]
&&\left\{\begin{array}{l}
\sqrt{\eta}\,\hat{E}_1(\rhovec,t) + \sqrt{1-\eta}\,\hat{E}_{{\rm vac}_1}(\rhovec,t) \\[.1in]
\sqrt{\eta}\, T(\rhovec) \,\hat{E}_2(\rhovec,t) + \sqrt{1-\eta |T(\rhovec)|^{2}}\,\hat{E}_{{\rm vac}_2}(\rhovec,t), \end{array}\right.
\end{eqnarray}
for $m=1,2$, where the $\{\hat{E}_{{\rm vac}_m}(\rhovec,t)\}$---which are needed to ensure commutator preservation---are in their vacuum states.  

The $\hat{C}(\rhovec_1)$ measurement yields an unbiased estimate of the ensemble-average equal-time photocurrent cross-correlation function
\begin{eqnarray}
\lefteqn{\langle \hat{C}(\rhovec_{1})\rangle  =  \langle \hat{\imath}_{1}(t) \hat{\imath}_{2}(t) \rangle 
 = q^2 \eta^{2} A_1} \nonumber \\[.12in]
 &\times& \int_{\mathcal{A}_{2}}\!d\rhovec\int\!du_1 \int\!du_2\, 
    h_{B}(t-u_{1}) h_{B}(t-u_{2}) |T(\rhovec)|^2\nonumber \\[.12in]
 &\times& \langle \hat{E}_{1}^{\dagger}(\rhovec_{1},u_{1}) \hat{E}_{2}^{\dagger}(\rhovec,u_{2}) \hat{E}_{1}(\rhovec_{1},u_{1}) \hat{E}_{2}(\rhovec, u_{2})  \rangle   \,, \label{IntCorr}
\end{eqnarray}
where we have approximated the integral over the pinhole detector's photosensitive region as the value of the integrand at $\rhovec_{1}$ times the photosensitive area $A_1$.
Evaluation of $\langle \hat{C}(\rhovec_{1})\rangle$, for the Gaussian-state sources we shall consider, can be accomplished along the lines established in \cite{ErkmenShapiro:GI}.  To find the ghost image signal-to-noise ratio at the point $\rhovec_1$, 
\begin{equation}
{\rm SNR} \equiv \frac{\langle \hat{C}(\rhovec_{1})\rangle^2}{\langle \Delta\hat{C}^2(\rhovec_{1})\rangle},
\end{equation}
where $\Delta \hat{C} (\rhovec) \equiv \hat{C}(\rhovec) - \langle \hat{C} (\rhovec) \rangle$, all that remains is to evaluate the variance term appearing in the denominator.

We have that the variance term obeys
\begin{eqnarray}
\lefteqn{\langle \Delta \hat{C}^{2}(\rhovec_{1}) \rangle = }  \nonumber \\[.1in]
&& \frac{1}{T_{I}^{2}} \int_{-T_{I}/2}^{T_{I}/2} \!dt \int_{-T_{I}/2}^{T_{I}/2}  \!du \, \langle \hat{\imath}_{1}(t) \hat{\imath}_{2}(t) \hat{\imath}_{1}(u) \hat{\imath}_{2}(u) \rangle \nonumber \\[.1in] 
&-& \langle \hat{C}(\rhovec_{1})\rangle^2. \label{var}
\end{eqnarray}
This expression reveals the primary challenge in evaluating the measurement variance: the fourth-moment of the photocurrents in the integrand is an eighth-order moment of the field operators. 
Fortunately, the moment-factoring theorem for Gaussian-state optical fields \cite{Mandel,Wozencraft}---which we used in \cite{ErkmenShapiro:GI} to find $\langle \hat{C}(\rhovec_{1})\rangle$---allows all field moments to be expressed in terms of second-order moments. Because this procedure is straightforward but tedious, we shall confine our discussion here to a detailed description of the simplification procedure, rather than a lengthy derivation. 

First, we express the integrand on the right-hand side of \eqref{var} in terms of the field-operator moments, as we have done in \eqref{IntCorr} for the mean. We then use the commutator relations \eqref{commutators:1}, \eqref{commutators:2} to put the integrand into normal order. This procedure yields the sum of four normally-ordered moments: one eighth-order moment, two sixth-order moments, and one fourth-order moment.  Next, the Gaussian-state moment-factoring theorem is applied to each term, replacing  higher-order moments with expressions that depend only on the second-order moments of the fields. Note that the nonzero terms in the moment-factored expression depend on whether the source of interest has nonzero phase-sensitive or phase-insensitive cross-correlation functions.
Finally, employing the coherence-separability of the correlation functions, the spatial and temporal integrals in each term are evaluated separately. It is relevant to note that many temporal integrals vanish due to our use of {\sc ac}-coupling, i.e., because $H_{B}(0)=0$.   Moreover, the symmetry properties of correlation functions can be used to group some nonzero terms so that the final variance expression is a sum of only eight terms. We now proceed with the details for each of the three sources under consideration.

\subsection{Thermal-State Light}

Lensless ghost imaging with thermal-state light usually derives its signal and reference sources from 50-50 beam splitting of a single zero-mean Gaussian-state beam possessing a phase-insensitive autocorrelation function but no phase-sensitive autocorrelation function, see, e.g., \cite{Scarcelli}.  Taking the post-splitter signal and reference fields to have the Gaussian-Schell model autocorrelations from \eqref{GS:PISauto}, it follows that these fields have the maximum phase-insensitive cross correlation, given by
\begin{eqnarray}
\lefteqn{\langle \hat{E}_{S}^{\dagger}(\rhovec_{1}, t_{1}) \hat{E}_{R}(\rhovec_{2}, t_{2}) \rangle =}
\nonumber \\[.1in]
&& \frac{2 P}{\pi a_{0}^2} e^{-(|\rhovec_{1}|^2+ |\rhovec_{2}|^2)/a_{0}^2 - |\rhovec_{2} - \rhovec_{1}|^2/2 \rho_{0}^2 } e^{-(t_2 - t_1)^2/2T_{0}^{2}},\hspace{.15in} \label{GS:PIScross}
\end{eqnarray}
and a vanishing phase-sensitive cross correlation, viz.,  $\langle \hat{E}_{S}(\rhovec_{1},t_{1}) \hat{E}_{R}(\rhovec_{2}, t_{2}) \rangle = 0$. 

Let us begin our thermal-state SNR analysis with near-field operation, wherein $k_{0} \rho_{0} a_{0}/2 L \gg 1$ prevails, with $k_0 \equiv \omega_0/c$ being the wave number associated with the center frequency $\omega_0$.  In this regime, the detection-plane correlation functions are approximately equal to those of the source. With all auto- and cross-correlation functions specified, evaluating the spatial and temporal integrals in the moment-factored variance expression is a straightforward exercise. For the spatial integrals, we assume that $a_{0}$ exceeds the transverse extent of the transmission mask by an amount sufficient to permit the approximation $e^{-|\rhovec|^{2}/a_{0}^{2}} |T(\rhovec)| \approx  |T(\rhovec)|$. For convenience, we define
\begin{equation}
A'_{T} \equiv \int\!d\rhovec \, |T(\rhovec)|^{4}\,,
\end{equation}
which we will regard as the effective area of the transmission mask.  Our $A_T'$ interpretation follows by analogy with the case of a binary ($|T(\rhovec)| \in \{0,1\}$) mask, for which $A'_T$ is the area over which $|T(\rhovec)| = 1$.  With this interpretation we have that $A'_{T}/\rho_{0}^{2}$ is the number of spatial resolution cells in the ghost image \cite{ErkmenShapiro:GI}. We also note that the small-pinhole approximation introduced in the previous section requires $\rho_{0}^{2} / A_1 \gg 1$ for its validity in near-field operation. Finally, we identify the two assumptions employed in evaluating the variance expression's temporal integrals:  $T_{I} \gg T_{0}$ and  $\Omega_{B} T_{I}\gg 1$.  Neither of these averaging-time conditions is at all surprising.  The former states that we must average over many source coherence times to form a high-quality ghost image.  The latter states that we must average over many photodetector response times to achieve this same purpose.  

Within the near-field regime---and subject to the assumptions given in the preceding paragraph---we will evaluate the ghost-imaging SNR behavior that prevails under narrowband and broadband illumination conditions.  A source state is said to be narrowband if $\Omega_{B} T_{0} \gg 1$, so that the coherence time of the impinging field state $T_0$ greatly exceeds the $\sim$$1/\Omega_B$ integration time of the photodetectors.  Conversely, a broadband source state is one that satisfies $\Omega_B T_0 \ll 1$, so that the source's coherence time is much shorter than the photodetector's integration time.  

For a narrowband source and near-field ghost imaging with thermal-state light when $A'_{T}/\rho_{0}^{2} \gg 30$ (e.g., the 2D image consists of $10 \times 10$ or more resolution cells), we find that the signal-to-noise ratio is 
\begin{eqnarray} 
\lefteqn{\text{SNR} = } \nonumber \\[.15in]
 && \hspace*{-.15in}\frac{|T(\rhovec_1)|^4T_{I}/T_{0}}{ \biggl [\frac{A'_{T}}{\sqrt{2\pi}\rho_{0}^{2}}  + \frac{|T(\rhovec_1)|^2}{\eta{\mathcal{I}}}+  \frac{4\pi\rho_{0}^{2}|T(\rhovec_1)|^4}{3A_1\eta{\mathcal{I}}}  + \frac{\sqrt{\pi}\Omega_{B} T_{0}\rho_0^2|T(\rhovec_1)|^2}{16 \sqrt{2}A_1\eta^2{\mathcal{I}}^2}  \biggr ]},\hspace{.25in} \label{SNR:ThNB}
\end{eqnarray}  
where ${\mathcal{I}} \equiv  PT_0\rho_0^2/a_0^2$ is the source brightness, i.e.,  the source's average number of photons per spatiotemporal mode.  As expected, this SNR expression grows linearly with increasing averaging time $T_I$, behavior that will be seen in all the cases we will consider in this paper.  More importantly, we can give physical interpretations to the terms in its noise denominator that dominate in low-brightness and high-brightness operation.

We have chosen to use quantum photodetection theory to derive all the SNR expressions in this paper.  However, as shown in \cite{ErkmenShapiro:GI}, quantitatively identical formulas follow from semiclassical photodetection theory when the signal-reference state is classical, i.e., when it has a proper $P$ representation.  The thermal state is classical.  It is thus appropriate to replace the photocurrent operators $\{\hat{\imath}_m(t)\}$ with classical photocurrents $\{i_m(t)\}$ that, owing to the assumed {\sc ac}-coupling, are zero-mean random processes comprised of a shot-noise component, arising from the discreteness of the electron charge, plus an excess noise component, which is proportional to the fluctuations in the photon flux illuminating the detector.  The variance contributions generated by these photocurrent components scale differently with increasing source brightness.  As a result, we can identify the left-most and right-most terms in the noise denominator of \eqref{SNR:ThNB}---which are the noise terms that dominate at high and low source brightness, respectively---as being normalized variance contributions coming from excess noise alone and from shot noise alone, while the middle terms arise from beating between excess noise and shot noise.   
Thus, as the source brightness grows without bound, the SNR from \eqref{SNR:ThNB} increases until it saturates at its maximum value, 
\begin{equation}
\text{SNR}_{\rm max} = \sqrt{2 \pi} \frac{T_{I} }{T_{0}} \frac{\rho_{0}^{2}}{A'_{T}}|T(\rhovec_1)|^4\,, \label{SNR:ThNBBri}
\end{equation}
which is limited by the excess-noise term \cite{footnote6}.  Roughly speaking, ${\rm SNR}_{\rm max}$ equals the number of source coherence times in the averaging interval divided by the number of spatial resolution cells in the image and multiplied by the square of the object's intensity transmission.  
Note that $\rho_{0}^{2}/ A'_{T}$ is the image contrast for {\sc dc}-coupled ghost-image formation in the near field with narrowband thermal-state light \cite{ErkmenShapiro:GI}.  Hence, the SNR of {\sc ac}-coupled, high-brightness, thermal-state ghost imaging is proportional to the image contrast realized using the same setup with {\sc dc}-coupling.

\begin{figure}
\centering
\subfigure[Narrowband.]{
		\includegraphics[width=2.75in]{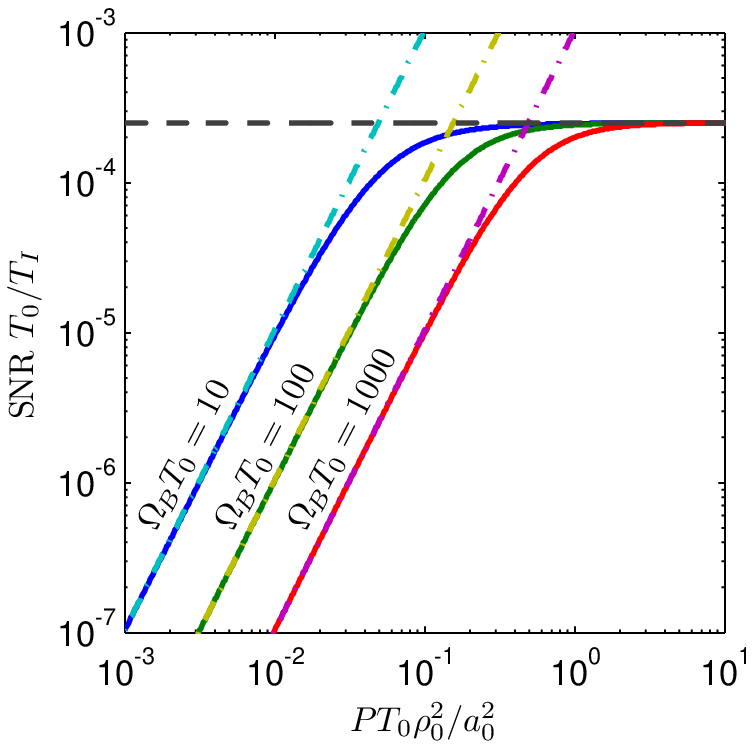}
		\label{Fig_SNR_ThNbNf} 
	}\\
\subfigure[Broadband.]{
		\includegraphics[width=2.75in]{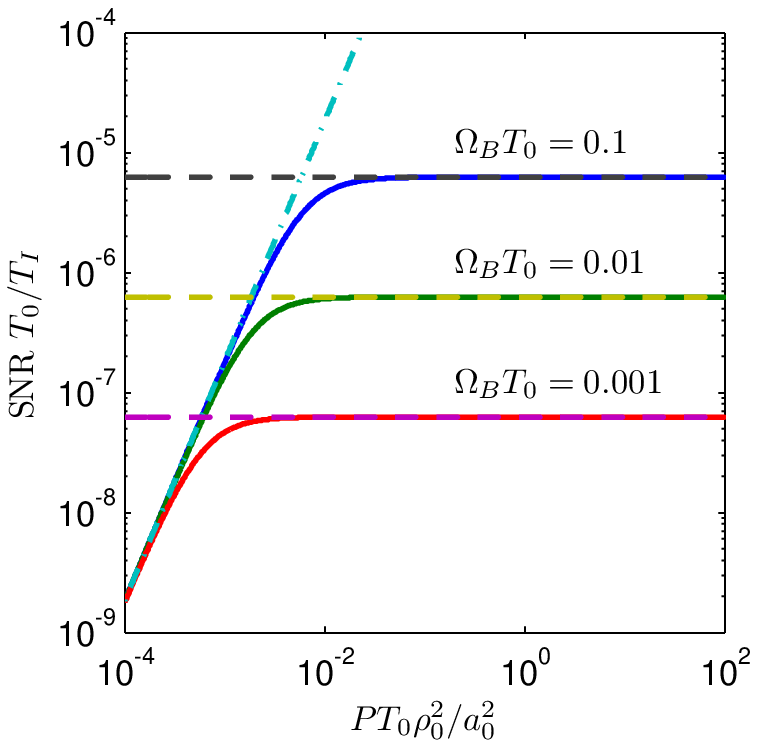}
		\label{Fig_SNR_ThBbNf}
	}
	\caption{(Color online)  Thermal-state ghost imaging SNR, normalized by $T_{I}/T_{0}$, plotted as a function of source brightness ${\mathcal{I}} \equiv P T_{0} \rho_{0}^{2}/a_{0}^{2}$, for $|T(\rhovec_1)| =1$, $A'_{T}/\rho_{0}^{2} = 10^4$, $\rho_{0}^{2}/A_1 = 10$ and $\eta = 0.9$. Various $\Omega_{B} T_{0}$ values are shown in the (a) narrowband and (b) broadband limits. Dash-dotted lines represent low-brightness asymptotes, and dashed lines correspond to high-brightness asymptotes. }
	\label{Fig_SNR_ThNf}
\end{figure}

At very low source brightness the ghost-image SNR obtained with a narrowband thermal-state source will be controlled by the shot-noise contribution to its noise denominator.  In this case \eqref{SNR:ThNB} reduces to 
\begin{equation}
\text{SNR} = \frac{16\sqrt{2} }{\sqrt{\pi}} \frac{T_{I}}{T_{0}} \frac{\eta P A_1}{ \Omega_{B} a_{0}^{2}}
\eta{\mathcal{I}}|T(\rhovec_1)|^2\,. \label{SNR:ThNBDim}
\end{equation}
In Fig.~\ref{Fig_SNR_ThNbNf} we have plotted the narrowband thermal-state ghost imaging SNR from \eqref{SNR:ThNB}---along with its high-brightness and low-brightness asymptotes from \eqref{SNR:ThNBBri} and \eqref{SNR:ThNBDim}---for several narrowband ghost imaging scenarios.

We now turn our attention to broadband sources, which satisfy $\Omega_{B} T_{0} \ll 1$. When $ A'_{T} /\rho_{0}^{2} \gg 12\, \Omega_{B} T_{0} $ holds, we find that
\begin{eqnarray}
\lefteqn{\text{SNR} = } \nonumber \\[.1in]
&& \hspace*{-.15in}\frac{|T(\rhovec_1)|^4T_{I}/T_{0}}{\biggl [  \frac{2\sqrt{2}A'_{T}}{\sqrt{\pi}\Omega_BT_0\rho_{0}^{2}} + \frac{2|T(\rhovec_1)|^2}{\sqrt{3}\eta{\mathcal{I}}} + \frac{8\pi \rho_0^2|T(\rhovec_1)|^4}{3\sqrt{3}A_1\eta{\mathcal{I}}} + \frac{\sqrt{\pi}\rho_{0}^{2}|T(\rhovec_1)|^2}{4A_1\eta^2{\mathcal{I}}^2}  \biggr ]},\hspace{.25in} \label{SNR:ThBB} 
\end{eqnarray}
where, once again, the left-most and right-most terms in the noise denominator are due to excess noise alone and shot noise alone.  Here too SNR increases with increasing source brightness until it reaches its maximum value, 
\begin{equation}
\text{SNR}_{\rm max} = \frac{\sqrt{\pi}}{2\sqrt{2}} \Omega_{B} T_{I} \frac{\rho_{0}^{2}}{A'_{T}}|T(\rhovec_1)|^4\,, \label{SNR:ThBBBri}
\end{equation}
where it is limited by excess noise alone.  Aside from insignificant numerical factors, the broadband SNR$_{\rm max}$ differs from the narrowband SNR$_{\rm max}$ only through replacement of $1/T_0$ from the narrowband expression with $\Omega_B$ in the broadband expression.  This replacement is to be expected.  In the narrowband case $T_I/T_0$ is the number of photocurrent coherence times that are  being averaged by the correlator.  This is because the narrowband condition $\Omega_B T_0 \gg 1$ ensures that the photon-flux fluctuations are not affected by the photodetector's baseband bandwidth limit.  However, under the broadband condition, $\Omega_BT_0 \ll 1$, the photocurrent fluctuations have a much longer ($\sim$$1/\Omega_B$) coherence time than that of the photon flux illuminating the detectors, so it is $\Omega_BT_I$ that appears in the broadband SNR$_{\rm max}$ formula.

At very low source brightness, the SNR of the broadband thermal-state ghost imager becomes limited by shot noise alone and is given by
\begin{equation}
\text{SNR} = \frac{4}{\sqrt{\pi}} \frac{T_{I}}{T_{0}} \frac{A_1}{\rho_{0}^{2}} \eta^2{\mathcal{I}}^2|T(\rhovec_1)|^2 \,. \label{SNR:ThBBDim}
\end{equation}
Figure~\ref{Fig_SNR_ThBbNf} shows several plots of broadband thermal-state ghost imaging SNR, together with its high-brightness and low-brightness asymptotes.

Thus far we have concentrated on the near-field SNR behavior with a thermal-state source. Our results, however, are easily converted to the far-field regime, in which $k_{0} \rho_{0} a_{0} /2 L \ll 1$. In order to obtain the far-field SNR we must first propagate the second-order correlation functions from the source plane to the detection planes. For Gaussian-Schell model correlation functions this transformation is a simple replacement of $a_{0}$ by $a_{L} \equiv 2 L / k_{0} \rho_{0} $,  and $\rho_{0}$ by $\rho_{L} \equiv 2 L / k_{0} a_{0}$, when quadratic phase factors that do not affect ghost-image formation are omitted \cite{footnote7}. Therefore, all of our near-field thermal-state SNR results can be converted to corresponding far-field results by making these parameter value changes. 

\subsection{Classically-correlated phase-sensitive light}

Let us now consider a source state that has the maximum phase-sensitive cross correlation permitted by classical physics, given the auto-correlation functions in \eqref{GS:PISauto}, but has no phase-insensitive cross-correlation, i.e.,
\begin{eqnarray}
\lefteqn{\langle \hat{E}_{S}(\rhovec_{1}, t_{1}) \hat{E}_{R}(\rhovec_{2}, t_{2}) \rangle = }
\nonumber \\[.1in]
&& \frac{2 P}{\pi a_{0}^2} e^{-(|\rhovec_{1}|^2+ |\rhovec_{2}|^2)/a_{0}^2 - |\rhovec_{2} - \rhovec_{1}|^2/2 \rho_{0}^2 } e^{-(t_2 - t_1)^2/2T_{0}^{2}}\,,\hspace{.15in} \label{GS:PScross}
\end{eqnarray}
and $\langle \hat{E}_{S}^{\dagger} (\rhovec_{1},t_{1}) \hat{E}_{R}(\rhovec_{2}, t_{2}) \rangle = 0$, where we have assumed the phase-sensitive cross correlation function is real valued~\cite{footnote8}.

In the near-field regime, which is now given by $k_{0} \rho_{0}^{2} / 2 L \gg 1$, we can follow the same assumptions stated in the previous section for thermal states and arrive at the \emph{same} near-field SNR expressions, i.e., \eqref{SNR:ThNB} and \eqref{SNR:ThBB} apply to narrowband and broadband near-field operation with this classically-correlated phase-sensitive source state. Furthermore, the high-brightness and low-brightness asymptotes  are as given by \eqref{SNR:ThNBBri} and \eqref{SNR:ThNBDim}, respectively, for the narrowband case, and by \eqref{SNR:ThBBBri} and \eqref{SNR:ThBBDim}, respectively, for the broadband case.  Moreover, the physical interpretations we provided for the thermal-state results continue to apply, without modification, for the classically-correlated phase-sensitive source.   

For phase-sensitive coherence, the far-field regime corresponds to $k_{0} a_{0}^{2} / 2 L \ll 1$. As stated in the previous section, the detection-plane (phase-insensitive)  auto-correlation functions are found by substituting $a_{L}$ for $a_{0}$ and $\rho_{L}$ for $\rho_{0}$. Propagating the source-plane phase-sensitive cross correlation given in \eqref{GS:PScross} involves the same substitutions, but in addition requires replacing $|\rhovec_{2} - \rhovec_{1}|^{2}$ by $|\rhovec_{2} + \rhovec_{1}|^{2}$, because the far-field ghost image formed with phase-sensitive light is inverted.    It follows that the the far-field SNR expressions are derived from the near-field SNR expressions by replacing the source-plane coherence radius $\rho_{0}$  and beam radius $a_{0}$ by their detection-plane counterparts, $\rho_{L}$ and $a_{L}$ respectively, and using $|T(-\rhovec_1)|$ in lieu of $|T(\rhovec_1)|$.  

\subsection{Maximally-entangled phase-sensitive light}

We continue to consider signal and reference beams in a zero-mean jointly-Gaussian state with no phase-insensitive cross correlation, but now we  take the phase-sensitive cross correlation to be the maximum permitted by \emph{quantum} physics. Because quantum ghost imaging experiments have used the signal and idler outputs from spontaneous parametric downconversion (SPDC) as the two source fields, we shall focus on this case here. The output field operators of SPDC can be expressed as~\cite{WongKimShapiro,Brambilla:SPDC,footnote9}
\begin{equation}
\hat{E}_{m}(\rhovec,t) = A(\rhovec) \hat{\mathcal{E}}_{m}(\rhovec,t) + \hat{\mathcal{L}}_{m}(\rhovec,t)\,,
\end{equation}
for $m=S,R$, where $|A(\rhovec)|\leq 1$ is an aperture function representing the finite transverse extent of the interaction medium, and $\hat{\mathcal{L}}_{m}(\rhovec,t)$ are auxiliary vacuum-state operators, so that the $\hat{E}_{m}(\rhovec,t)$ satisfy the free-space field commutator relations. The operator-valued Fourier transforms of $\{ \hat{\mathcal{E}}_{m}(\rhovec,t), m=S,R\}$, denoted by $\{\hat{A}_{m}(\kvec, \Omega), m=S,R \}$, are given by a two-field Bogoliubov transformation of vacuum-state input field operators, $\hat{a}_{m}(\kvec, \Omega)$, i.e., 
\begin{eqnarray}
\lefteqn{\hat{A}_{S}(\kvec, \Omega) =} \nonumber \\ &&
 \mu(\kvec,\Omega) \hat{a}_{S}(\kvec,\Omega) \!+\! \nu(\kvec, \Omega) \hat{a}_{R}^{\dagger}(-\kvec,-\Omega)\\
\lefteqn{\hat{A}_{R}(-\kvec, -\Omega) = } \nonumber \\ &&
 \mu(\kvec,\Omega) \hat{a}_{R}(-\kvec,-\Omega) \!+\! \nu(\kvec, \Omega) \hat{a}_{S}^{\dagger}(\kvec,\Omega).
\end{eqnarray}
Here $\nu(\kvec,\Omega) \in \mathbb{R} $ and $\mu(\kvec,\Omega) \equiv 1 + i \nu(\kvec,\Omega)$ are the canonical transformation coefficients. In accordance with the Gaussian-Schell model treatment introduced earlier, we set~\cite{footnote10}
\begin{equation}
\nu(\kvec, \Omega) = 2 (2 \pi)^{1/4} \sqrt{\frac{P T_{0} \rho_{0}^{2}}{a_{0}^{2}}} e^{-\rho_{0}^{2} |\kvec|^{2}/4 - T_{0}^{2} \Omega^{2}/4}\,,
\end{equation}
and
\begin{equation}
A(\rhovec) = \exp\{-|\rhovec|^{2} / a_{0}^{2}\}\,,
\end{equation}
such that the $\hat{E}_{m}(\rhovec,t)$, for $m = S,R$, are in a zero-mean jointly Gaussian state, with phase-insensitive autocorrelation functions given by \eqref{GS:PISauto}, and the maximum permissible phase-sensitive cross correlation function, 
\begin{eqnarray}
\nonumber \lefteqn{\langle \hat{E}_S(\rhovec_1,t_1)\hat{E}_R(\rhovec_2,t_2)\rangle = \frac{2 P}{\pi a_{0}^2} e^{-(|\rhovec_{1}|^2+ |\rhovec_{2}|^2)/a_{0}^2 }} \\[0.1in] 
 \nonumber
&\times& \hspace*{-.05in}\biggl [ e^{- |\rhovec_{2} - \rhovec_{1}|^2/2 \rho_{0}^2 } e^{-(t_2 - t_1)^2/2T_{0}^{2}}  \nonumber \\[.1in] 
&+&  \hspace*{-.05in}i (2/\pi)^{1/4} \sqrt{\frac{a_{0}^{2}}{ P T_{0} \rho_{0}^2}}  e^{- |\rhovec_{2} - \rhovec_{1}|^2/\rho_{0}^{2} } e^{-(t_2 - t_1)^2/T_{0}^{2}} \biggr ]\!.
 \label{GS:PScrossQ}
\end{eqnarray}
All other second-order moments, i.e., the phase-sensitive autocorrelation functions and the phase-insensitive cross-correlation function, are zero. It is worthwhile to point out that when the source brightness $\mathcal{I} \equiv P T_{0} \rho_{0}^{2} / a_{0}^{2} \gg 1$, the first term in the square brackets dominates, and \eqref{GS:PScrossQ} approaches the classical phase-sensitive cross correlation given in \eqref{GS:PScross}. However, when $\mathcal{I} \ll 1$, the second term is much larger than the first, resulting in a much stronger phase-sensitive cross correlation than permitted in a classical state. If the brightness is lowered to the limit in which there is on average much less than one photon in the signal and idler beams, the output of the SPDC can be approximated as a dominant vacuum component plus a weak pair of entangled photons, viz., the biphoton state~\cite{WongKimShapiro,ErkmenShapiro:GI}.

\begin{figure}
\centering
\subfigure[Narrowband.]{
		\includegraphics[width=2.75in]{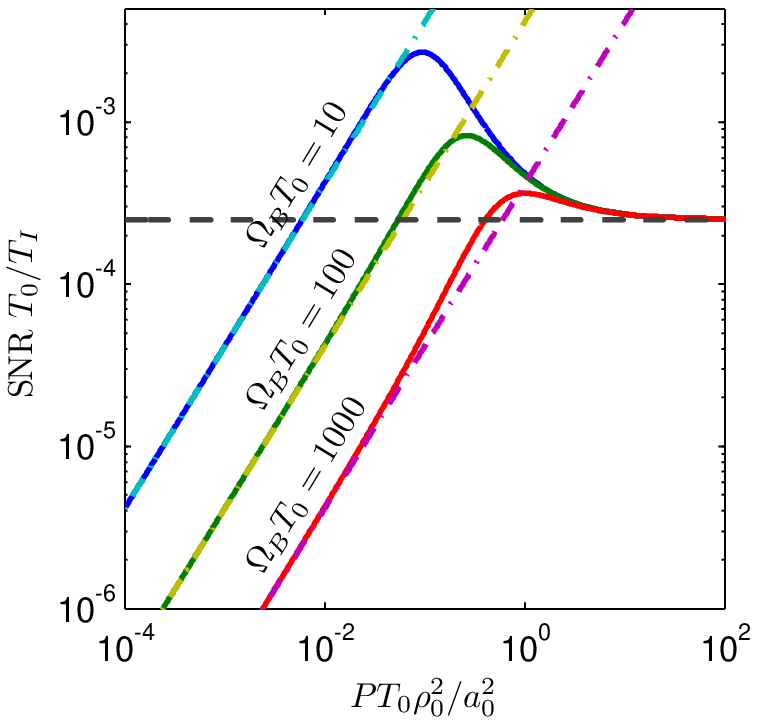}
		\label{Fig_SNR_BiNbNf} 
	}\\
\subfigure[Broadband.]{
		\includegraphics[width=2.75in]{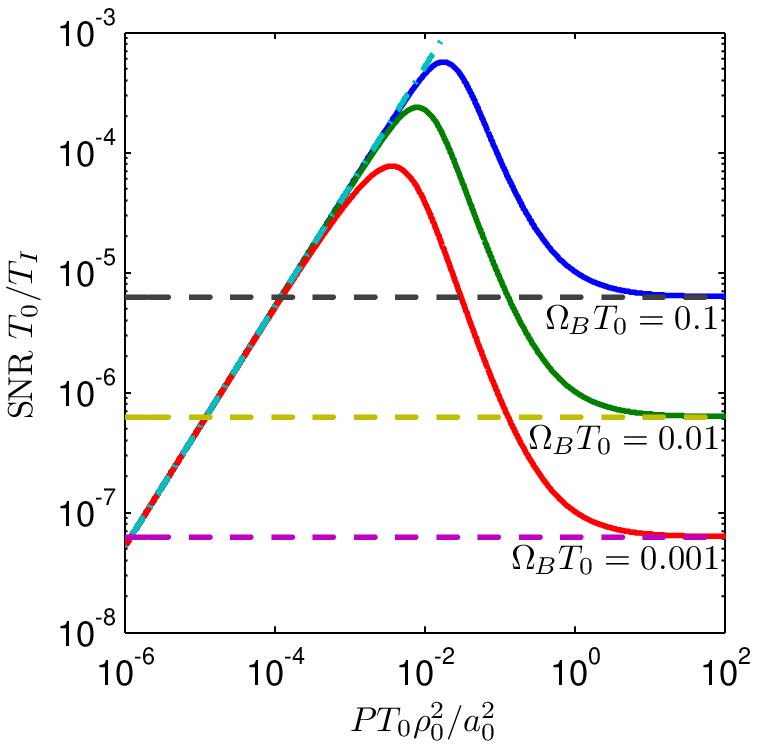}
		\label{Fig_SNR_BiBbNf}
	}
	\caption{(Color online) Nonclassical phase-sensitive Gaussian-state ghost-imaging SNR, normalized by $T_{I}/T_{0}$, plotted versus source brightness ${\mathcal{I}} \equiv P T_{0} \rho_{0}^{2}/a_{0}^{2}$  for a near-field configuration ($k_{0} \rho_0^{2}/ 2 L \gg 1$) with $|T(\rhovec_1)| = 1$, $A'_{T}/\rho_{0}^{2} = 10^4$, $\rho_{0}^{2}/A_1 = 10$, and $\eta = 0.9$. Various $\Omega_{B} T_{0}$ values are shown in the (a) narrowband and (b) broadband limits. Dash-dotted lines represent low-brightness asymptotes and dashed lines correspond to high-brightness asymptotes.}
	\label{Fig_SNR_BiNf}
\end{figure}

To evaluate the ghost-image SNR in the near-field regime ($k_{0} \rho_{0}^{2} / 2 L \gg 1$) we utilize the same approximations we have used for classical phase-sensitive light, now with the cross-correlation function from \eqref{GS:PScrossQ} employed in lieu of \eqref{GS:PScross} when integral expressions are explicitly evaluated.   In the narrowband limit, $\Omega_{B} T_{0} \gg 1$, we find that
\begin{widetext}
\begin{equation}
\text{SNR} = \frac{ \Bigl (1+1/\sqrt{2 \pi} \mathcal{I} \Bigr )^{2} T_{I}/T_{0}}{ \Biggl [ \frac{A'_{T}}{\sqrt{2\pi} |T(\rhovec_1)|^4 \rho_{0}^{2}}  + \frac{1}{\mathcal{I}} \Bigl ( \frac{1}{\eta |T(\rhovec_{1})|^{2}} + \frac{4 \pi \rho_{0}^{2}}{3 A_{1} \eta }  \Bigr )  + \frac{\sqrt{\pi} \rho_{0}^{2} \Omega_{B} T_{0} }{16\sqrt{2}  A_{1} |T(\rhovec_1)|^{2} \eta^{2} \mathcal{I}^{2}} \Bigl  (1 \!+\! \frac{1}{\sqrt{2\pi} \mathcal{I}} \Bigr )\Biggr ]}.\label{SNR:BiNB}
\end{equation}
\end{widetext}
This SNR expression captures the full quantum-to-classical transition seen in ghost imaging with maximally-entangled phase-sensitive light (the output fields from SPDC), as the source brightness $\mathcal{I}$ is increased. When the source is bright, i.e., $\mathcal{I} \gg 1$, the first terms in the numerator and the denominator of \eqref{SNR:BiNB} are dominant, yielding the same SNR as that obtained with narrowband, bright classical maximally-correlated (thermal or phase-sensitive) light, i.e., \begin{equation}
\text{SNR} = \sqrt{2 \pi} \frac{T_{I} }{T_{0}} \frac{\rho_{0}^{2}}{A'_{T}}|T(\rhovec_1)|^4\,. \label{SNR:BiNBBri}
\end{equation}
For dim-source  ($\mathcal{I} \ll 1$) ghost imaging the second term in the numerator of \eqref{SNR:BiNB} dominates the first, and when $|T(\rhovec_1)| \sim 1$ the last term in the noise denominator is the most significant, yielding an SNR that is linear in photon flux,
\begin{equation}
\text{SNR} = \frac{16}{\pi}  \frac{T_{I}}{T_{0}} \frac{\eta^{2} P  A_1|T(\rhovec_1)|^2}{\Omega_{B} a_{0}^{2}} \,. \label{SNR:BiNBCC}
\end{equation}
For example, even with the generous parameters values $P T_{0} \rho_{0}^{2} / a_{0}^{2} = 0.01$, $\rho_{0}^{2} / A_1 = 10$, and $\Omega_{B} T_{0} =10$, it is necessary to have $10^{7}$ resolution cells before \eqref{SNR:BiNB} deviates appreciably from the linear dependence on the photon flux $P$ for $|T(\rhovec_1)| \sim 1$. In this regime the SNR is limited by the very low number of photon pairs detected over a detector integration time.  The SNR achieved with narrowband maximally-entangled phase-sensitive light is plotted in Fig.~\ref{Fig_SNR_BiNbNf} for several $\Omega_{B} T_{0}$ values. The plots verify the linear low-brightness regime and the high-brightness saturation towards the classical asymptote. However, as shown in the plotted curves, the SNR can exceed the bright-source asymptote. When this occurs, there is a finite source brightness that yields the maximum SNR, and increasing $\mathcal{I}$ beyond this threshold will \emph{decrease} the SNR with increasing photon flux. For a given set of parameters this source-brightness threshold can be found easily by solving for the roots of a third-order polynomial, which yields the critical points of \eqref{SNR:BiNB}. Although closed-form solutions for these roots exist, the expressions are too cumbersome to pursue further in this paper.

If the low-brightness condition of the source ($\mathcal{I} \ll 1$) is augmented with the \emph{low-flux} condition
\begin{equation}
\frac{\eta P A'_{T} }{\Omega_{B} a_{0}^{2}} \ll1 \,,\label{CCRegime}
\end{equation}
then the average number of photons per integration time impinging on either detector becomes much less than unity. It follows that the photodetectors can be replaced with non-photon-resolving photodetectors without appreciable loss in imaging functionality, thereby rendering the Fig.~\ref{GI:propagation} ghost imaging configuration equivalent to biphoton-state ghost imaging with coincidence-counting circuitry (instead of photocurrent correlation). Thus, narrowband biphoton-state ghost imaging is also governed by the linear photon-flux SNR formula from \eqref{SNR:BiNBCC} for $|T(\rhovec_1)| \sim 1$.

Shifting our attention to the broadband ($\Omega_{B} T_{0} \ll 1$) limit, we arrive at
\begin{widetext}
\begin{equation}
\text{SNR} = \frac{\Bigl ( 1 + 1/2\sqrt{\pi} \mathcal{I} \Bigr )^{2} T_{I}/T_{0}}{ \Biggl [ \frac{\sqrt{8} A'_{T}}{\sqrt{\pi} \Omega_{B} T_{0} \rho_{0}^{2} |T(\rhovec_1)|^{4}} + \frac{D_1}{\mathcal{I}}   + \frac{1}{2 \sqrt{\pi} \mathcal{I}^{2}} \Bigl(\frac{\sqrt{2}}{\Omega_{B} T_{0}} + \frac{\pi \rho_{0}^{2}}{\eta A_{1}  }  \bigl (1 + \frac{1}{2 \eta |T(\rhovec_1)|^{2}} \bigr ) \Bigr) + \frac{\rho_{0}^{2}}{8 \eta^{2} A_{1} |T(\rhovec_1)|^{2} \mathcal{I}^{3} } \Biggr ]}\,, \label{SNR:BiBB}
\end{equation}
\end{widetext}
where
\begin{equation}
D_1  \equiv \frac{2}{\sqrt{3} \eta |T(\rhovec_{1})|^{2}} + \frac{\sqrt{8}}{\Omega_{B} T_{0}} + \frac{ 8 \pi \rho_{0}^{2}}{3\sqrt{3} \eta A_{1}}.
\end{equation} 
When $\mathcal{I} \gg 1$, the first terms in the numerator and  denominator are dominant. Consequently, the SNR approaches the SNR of classical ghost imaging with a bright (phase-insensitive or phase-sensitive) maximally-correlated broadband source, which is given in \eqref{SNR:ThBBBri}.

On the other hand, if the low-brightness condition, $\mathcal{I} \ll 1$, and the low-flux condition,  as given in \eqref{CCRegime},  are both satisfied, the right-most term in the noise denominator becomes dominant, for $|T(\rhovec_1)| \sim 1$, yielding
\begin{equation}
\text{SNR} = \frac{2}{\pi}  \Omega_{B} T_{I} \frac{\eta^{2} P A_1|T(\rhovec_1)|^2}{\Omega_{B} a_{0}^{2}} \label{SNR:BiBBCC}
\end{equation}
for the broadband biphoton-state SNR expression. As in the narrowband case, the SNR in this regime suffers from the paucity of photon pairs detected within the photodetector's $\sim$$1/\Omega_B$ integration time, which is compensated by averaging the photocurrent product over many temporal coherence bins, i.e., employing $T_{I}\Omega_{B} \gg 1$. Figure~\ref{Fig_SNR_BiBbNf} shows plots of \eqref{SNR:BiBB} for several values of $\Omega_{B} T_{0}$. It is seen that the linear photon-flux dependence of the SNR extends well beyond the low-flux regime. Furthermore, the SNR achieved in the low-brightness regime (i.e., \eqref{SNR:BiBBCC}) can exceed the bright classical source asymptote given in \eqref{SNR:ThBBBri}. Similar to what we found for narrowband operation, the broadband SNR has a well-defined maximum achieved at finite source brightness. Increasing the brightness beyond this optimal value reduces the SNR, which ultimately converges to the SNR attained with classical sources. All three of these SNR regimes are clearly identifiable in the Fig.~\ref{Fig_SNR_BiBbNf} SNR plots. For example, for $\Omega_{B} T_{0} = 10^{-2}$, the SNR is linear in photon flux until ${\mathcal{I}} \approx 10^{-3} $, its maximum occurs at $\mathcal{I}\approx 10^{-2}$, then the SNR decreases with increasing $\mathcal{I}$ until at $\mathcal{I} \approx 10$ it converges to the bright-source asymptote.

\begin{figure}
\centering
\subfigure[Narrowband.]{
		\includegraphics[width=2.75in]{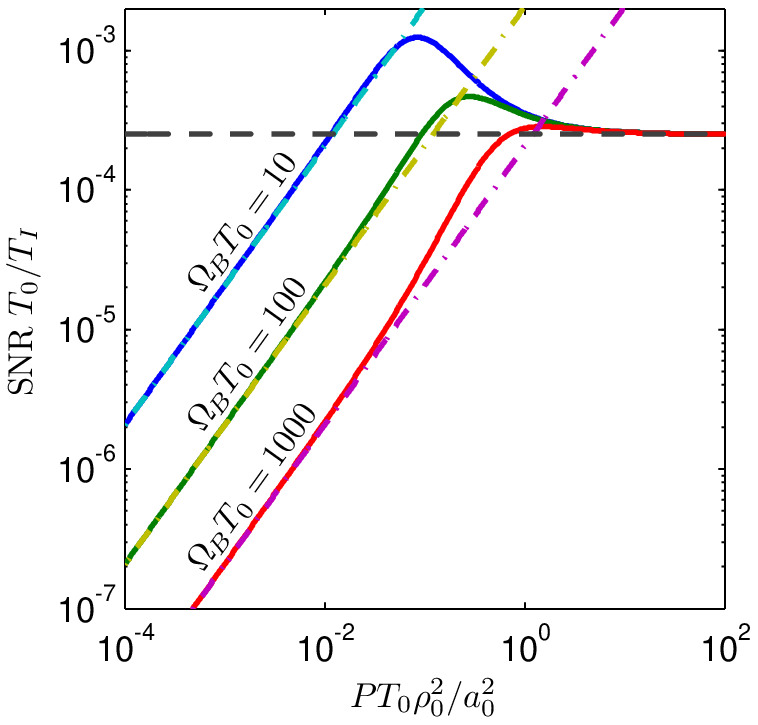}
		\label{Fig_SNR_BiNbFf} 
	}\\
\subfigure[Broadband.]{
		\includegraphics[width=2.75in]{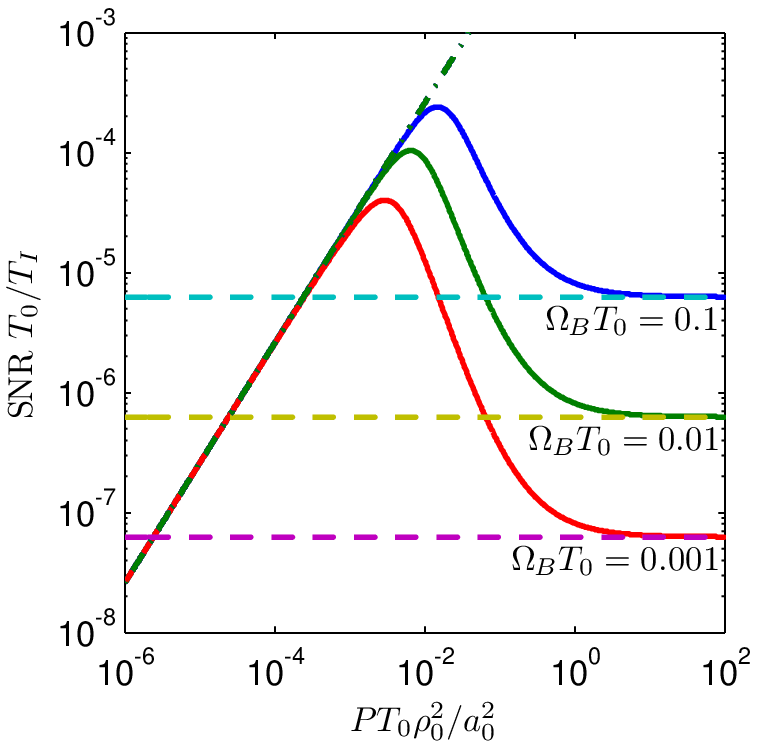}
		\label{Fig_SNR_BiBbFf}
	}
	\caption{(Color online) Nonclassical phase-sensitive Gaussian-state ghost imaging SNR, normalized by $T_{I}/T_{0}$, plotted versus source brightness ${\mathcal{I}} \equiv P T_{0} \rho_{0}^{2}/a_{0}^{2}$ for a far-field configuration ($k_{0} a_{0}^{2}/2L\ll 1$) with $|T(-\rhovec_1)|=1$, $A'_{T}/\rho_{0}^{2} = 10^4$, $\rho_{0}^{2}/A_1 = 10$ and $\eta = 0.9$. Various $\Omega_{B} T_{0}$ values are shown in the (a) narrowband and (b) broadband limits. Dash-dotted lines represent low-brightness asymptotes and dashed lines correspond to high-brightness asymptotes.}
	\label{Fig_SNR_BiFf}
\end{figure}

In the far field ($k_{0} a_{0}^{2}/2L  \ll 1$) the source-plane phase-sensitive cross correlation in \eqref{GS:PScrossQ} must be propagated to the detection planes \cite{footnote7} before the SNR can be evaluated. For narrowband sources ($\Omega_{B} T_{0} \gg 1$) we get
\begin{widetext}
\begin{equation}
\text{SNR} = 
 \frac{ \Bigl (1+1/\sqrt{8 \pi} \mathcal{I} \Bigr )^{2} T_{I}/T_{0}}{ \Biggl [ \frac{A'_{T}}{\sqrt{2\pi} |T(-\rhovec_1)|^4 \rho_{L}^{2}}  + \frac{1}{\mathcal{I}} \Bigl (\frac{1}{\eta |T(-\rhovec_{1})|^{2}} + \frac{4 \pi \rho_{L}^{2}}{3 A_{1} \eta } \Bigr) + \frac{\sqrt{\pi} \rho_{L}^{2} \Omega_{B} T_{0} }{16\sqrt{2}  A_{1}  |T(-\rhovec_1)|^{2} \eta^{2} \mathcal{I}^{2}} \Bigl  (1 \!+\! \frac{1}{\sqrt{8\pi} \mathcal{I}} \Bigr )\Biggr ]}. \label{SNR:BiNBFF}
\end{equation}
\end{widetext}
This result simplifies to 
\begin{equation}
\text{SNR} = \frac{8}{\pi} \frac{T_{I}}{T_{0}} \frac{\eta^{2} P A_1|T(-\rhovec_1)|^2 }{\Omega_{B} a_{L}^{2}}
\end{equation}
in the low-flux (biphoton-state) limit, defined by \eqref{CCRegime}, with $|T(-\rhovec_1)| \sim 1$, showing that narrowband biphoton-state ghost imaging SNR is limited by the number of photon pairs detected within the detectors' integration time.  On the other hand, when $\mathcal{I} \gg 1$, the SNR converges to
\begin{equation}
\text{SNR} = \sqrt{2\pi} \frac{T_{I}}{T_{0}} \frac{\rho_{L}^{2}}{ A'_{T}} |T(-\rhovec_1)|^{4}\,,
\end{equation}
which, as explained in the previous subsection, is equal to the bright-source SNR asymptote for far-field ghost imaging with narrowband classical (maximally-correlated) phase-sensitive light. Similar to the near-field scenarios, the SNR can have a maximum at finite source brightness. The full behavior of the far-field biphoton SNR in the narrowband regime is shown in Fig.~\ref{Fig_SNR_BiNbFf}.

The SNR in the broadband case ($ \Omega_{B} T_{0} \ll 1$), on the other hand, is given by
\begin{widetext}
\begin{equation}
\text{SNR} = \frac{\Bigl ( 1 + 1/4 \sqrt{\pi} \mathcal{I} \Bigr )^{2} T_{I}/T_{0}}{ \Biggl [ \frac{\sqrt{8} A'_{T}}{\sqrt{\pi} \Omega_{B} T_{0} \rho_{L}^{2} |T(-\rhovec_1)|^{4}} + \frac{1}{\mathcal{I}} \Bigl(\frac{2}{\sqrt{3} \eta |T(-\rhovec_{1})|^{2}}+ \frac{\sqrt{2}}{\Omega_{B} T_{0}} + \frac{ 8\pi \rho_{L}^{2}}{3\sqrt{3} \eta A_{1}} \Bigr)   + \frac{D_2}{8 \sqrt{\pi}  \mathcal{I}^{2}}  + \frac{\rho_{L}^{2}}{16 \eta^{2} A_{1} |T(-\rhovec_1)|^{2} \mathcal{I}^{3} } \Biggr ]}\,, \label{SNR:BiBBFF}
\end{equation}
\end{widetext}
where
\begin{equation}
D_2 \equiv \frac{\sqrt{2}}{\Omega_{B}T_{0}} + \frac{8 \pi \rho_{L}^{2}}{3 \eta A_{1}  } \Bigl (1 + \frac{3}{4 \eta |T(-\rhovec_1)|^{2}}\Bigr ),
\end{equation}
which simplifies in the biphoton limit to
\begin{equation}
\text{SNR} = \frac{1}{\pi} \Omega_{B} T_{I} \frac{\eta^{2} P A_1|T(-\rhovec_1)|^2}{\Omega_{B} a_{L}^{2}}, \label{SNR:BiBBCCFF}
\end{equation}
for $|T(-\rhovec_1)| \sim 1$.  
Hence, broadband biphoton-state ghost-imaging SNR is limited by the number of photon pairs detected within  the detectors' $\sim\,1/\Omega_B$ integration time.  The $\mathcal{I} \gg 1$ SNR simplifies to
\begin{equation}
\text{SNR} = \sqrt{\frac{\pi}{8}} \Omega_{B} T_{I} \frac{\rho_{L}^{2}}{A'_{T}} |T(-\rhovec_1)|^{4}\,,
\end{equation}
which is identical to the bright-source SNR asymptote for far-field ghost imaging with broadband classical (maximally-correlated) phase-sensitive light. The far-field biphoton SNR in the broadband regime is shown in Fig.~\ref{Fig_SNR_BiBbFf} for several $\Omega_{B} T_{0}$ values.

\section{Image Acquisition Times}
\label{IntTimes}

All of our SNR expressions are proportional to the cross-correlation averaging time $T_{I}$.  Consequently it is meaningful to compare the averaging times required to achieve a desired SNR value with different ghost imagers. Let us first assume that all parameters except photon flux are equal in the configurations of interest. Because the classical sources (the thermal state and the classical phase-sensitive state) yield identical SNRs, we shall use $P^{(c)}$ to denote their photon fluxes, reserving  $P^{(q)}$ for the photon flux of  the maximally-entangled (nonclassical) state.  Likewise, the averaging time for the classical-state ghost imagers to achieve the desired SNR will be denoted by $T_{I}^{(c)}$, while that for the entangled-state source will be designated $T_{I}^{(q)}$. 

Then, in the near field and with narrowband sources we obtain \cite{footnote11}
\begin{equation} 
\frac{T_{I}^{(q)}}{T_{I}^{(c)}} = \frac{\pi \sqrt{\pi}}{8\sqrt{2}} \frac{\Omega_{B} a_{0}^{2}}{\eta^{2} P^{(q)}  A'_{T}} \frac{\rho_{0}^{2}}{A_1}|T(\rhovec_1)|^2 \,, \label{NbRatio}
\end{equation}
where the classical-state sources are assumed to be bright enough to achieve the saturation SNR in \eqref{SNR:ThNBBri}, but the quantum source is limited to low brightness so that the nonclassical signature of the source prevails.  In the biphoton-state (low-flux) limit with $|T(\rhovec_1)| \sim 1$, \eqref{NbRatio} implies
$T_{I}^{(q)} \gg T_{I}^{(c)}$,
i.e., the cross-correlation averaging time required for narrowband biphoton-state ghost imagers to achieve a desired SNR value is much longer than that for bright classical-state ghost imagers to do so, given that all other system parameters are equal.

In the near field using broadband sources we obtain
\begin{equation} 
\frac{T_{I}^{(q)}}{T_{I}^{(c)}} = \frac{\pi \sqrt{\pi}}{4 \sqrt{2} } \frac{\Omega_{B} a_{0}^{2}}{\eta^{2} P^{(q)} A'_{T}} \frac{\rho_{0}^{2}}{A_1}|T(\rhovec_1)|^2\,. \label{IntTime:BB}
\end{equation} 
Thus, when the quantum ghost imager utilizes a low-flux (biphoton-state) source and $|T(\rhovec_1)| \sim 1$, we  find that
$T_{I}^{(q)} \gg T_{I}^{(c)}$
prevails, i.e., once again the cross-correlation averaging time required for ghost imaging with broadband bright classical-state light is significantly shorter than that for biphoton-state ghost imaging, given equal system parameters. Nevertheless, in a very high-resolution ghost imaging configuration, high illumination flux ($P^{(q)} A'_{T}/ \Omega_{B} a_{0}^{2} \gg 1 $) may be achievable with low-brightness ($P^{(q)} T_{0} \rho_{0}^{2}/a_{0}^{2} \ll 1 $) maximally-entangled phase-sensitive Gaussian-state light. In this case \eqref{IntTime:BB} implies that the averaging time for the nonclassical-state ghost imager can be shorter than that for the classical-state ghost imager \cite{footnote12}. For example, $|T(\rhovec_1)| = 1$, $A'_{T}/\rho_{0}^{2} = 10^{4}$, $\Omega_{B} T_{0} = 10^{-2}$ $P^{(q)} T_{0} \rho_{0}^{2} /a_{0}^{2} = 10^{-3}$, and $\rho_0^2/A_1 = 10$  will yield
$T_{I}^{(q)} \approx T_{I}^{(c)}/100\eta^{2}$.

Finally, we compare ghost imaging with a broadband biphoton-state (low-brightness and low-flux) to that with a bright narrowband classical state. Denoting the parameters specific to the classical and quantum sources with the superscripts $(c)$ and  $(q)$, we obtain
\begin{equation}
\frac{T_{I}^{(q)}}{T_{I}^{(c)}} = \frac{\pi \sqrt{\pi}}{\sqrt{2}} \frac{\Omega_{B}^{(q)} a_{0}^{2}}{\eta^{2} P^{(q)} A'_{T}} \frac{\rho_{0}^{2}}{A_1} \frac{|T(\rhovec_1)|^2}{\Omega_{B}^{(q)} T_{0}^{(c)}}\,.
\end{equation} 
Because the last factor on the right is typically less than unity, for $|T(\rhovec_1)| \sim 1$, whereas the remaining factors are greater than unity, the cross-correlation averaging time may be favor either source state.  As an example, consider $|T(\rhovec_1)| = 1$, $A'_{T}/\rho_{0}^{2} = 10^{4}$, $P^{(q)} T_{0}^{(q)} \rho_{0}^{2} /a_{0}^{2} = 10^{-6}$, $\rho_{0}^{2} / A_1 = 10$. Then a  biphoton-state source with 1\,THz bandwidth and a 1\,MHz thermal-state source will result in
$T_{I}^{(q)} \approx 4\times10^{-3} T_{I}^{(c)}/\eta^{2}$,
which shows that the biphoton-state imager enjoys an enormous advantage in averaging time as compared to the classical-state imager when the quantum efficiency is not unduly low.  

So, ghost imaging with bright classical thermal or phase-sensitive states  affords a shorter averaging time to reach a desired SNR value than does a biphoton-state ghost imager, given all of the remaining parameters governing the two ghost imaging systems are equal and $|T(\rhovec_1)| \sim 1$. However, ghost imaging with low-brightness, but high-flux quantum sources may achieve shorter averaging times than bright classical sources, if high quantum efficiency photon-number resolving photodetectors are employed. Finally, if ghost imaging with a broadband biphoton state is compared to that with a narrowband bright classical state, the integration time may favor either source, depending on the ratio of the achievable source bandwidths, number of resolution cells in the image, and the biphoton-state source brightness.

\section{Discussion}
\label{discussion}

We have presented a detailed SNR analysis for three Gaussian-state ghost imaging configurations.   Two used classical-state light, specifically a joint signal-reference field state that has either the maximum phase-insensitive or the maximum phase-sensitive cross correlation consistent with having a proper $P$ representation.  The third used nonclassical light, in particular an entangled signal-reference field state obtained from SPDC, with the maximum phase-sensitive cross correlation permitted by quantum mechanics.  Our analysis concentrated on the narrowband and broadband limits in both the near field and the far field. Because the conclusions from our analysis apply, in identical ways, to both the near-field and far-field regimes, we shall omit references to these regimes in what follows.

We found that classical-state ghost imager SNRs saturate---with increasing source brightness---to  maximum values that are inversely proportional to the number of resolution cells on the imaged object. In this high-brightness limit the SNR is thus proportional to the contrast achieved in {\sc dc}-coupled operation. The contrast can be improved by removing the featureless background via {\sc ac}-coupled photodetectors or background subtraction, but SNR improvements, at high source brightness, require  increasing the cross-correlation averaging time.

Biphoton-state ghost imagers were shown to have SNRs that are typically proportional to their low photon flux. Hence, for such imagers, increasing the photon flux of the source is of prime importance. However, the SNR gain derived from such increases is not unbounded. Our analysis revealed that the SNR realized with a nonclassical-state source with low brightness but high flux typically has a well-defined maximum, after which increasing flux \emph{reduces} the SNR. The inverse relation between SNR and photon flux may seem counterintuitive, but it is consistent with the fact that the SNR must approach the bright classical-state SNR as the source brightness increases beyond unity. This classical-state limit, however, is in general \emph{lower} than the maximum SNR achieved by the nonclassical-state source in the low-brightness regime. Hence, in these cases, the SNR achieved with nonclassical phase-sensitive light must have a decreasing trend as source brightness increases without bound.

To assess the performance achieved by different sources we compared their image acquisition times, i.e., the cross-correlation averaging times needed to achieve a predefined target value for SNR. We showed that with equal bandwidth sources, and all system parameters being equal unless otherwise noted, bright classical-state ghost imagers typically reach the desired SNR value with a much shorter averaging time than that needed by a biphoton-state ghost imager. Therefore, although the biphoton-state yields images with high contrast even in {\sc dc}-coupled operation, because the biphoton imager is photon starved the total time duration it requires to accumulate the ghost image far exceeds that necessary with a bright classical-state source. Nevertheless, we saw that there is a broadband, low-brightness, high-flux regime of nonclassical phase-sensitive light that may get by with much shorter cross-correlation averaging times than those needed by bright classical-state imagers. The notable drawback to reaping this quantum advantage, however, is the necessity for high quantum efficiency photon-number resolving detectors.

In conclusion, Gaussian-state analysis provides a robust and versatile framework for answering some of the most fundamental questions associated with developing practical ghost imagers for remote sensing applications. In this paper, we have used this framework to study the SNR behavior of ghost imagers with source states that encompass those that have been used in proof-of-principle ghost imaging experiments. Our analysis unambiguously identifies the key parameters that limit SNR  behavior.  For high-brightness classical-state ghost imaging it is the number of resolution cells in the image, whereas for the biphoton-state case it is the low photon flux of the source.

\section*{ACKNOWLEDGMENTS}
This work was supported by the DARPA Quantum Sensors Program, the U.S. Army Research Office MURI Grant No. W911NF-05-1-0197, and the W. M. Keck Foundation Center for Extreme Quantum Information Theory.


\begin{thebibliography}{10}

\bibitem{Gatti:one} A. Gatti, E. Brambilla, and L. A. Lugiato, Phys. Rev. Lett. {\bf 90,} 133603 (2003); 

\bibitem{Gatti:two} A. Gatti, E. Brambilla, M. Bache, and L. A. Lugiato, Phys. Rev. Lett. {\bf 93,} 093602 (2004); 

\bibitem{Gatti:three} A. Gatti, E. Brambilla, M. Bache, and L. A. Lugiato, Phys. Rev. A {\bf 70,} 013802 (2004); 

\bibitem{Cai:one} Y. Cai and S.-Y. Zhu, Opt. Lett. E {\bf 29,} 2716 (2004); 

\bibitem{Cai:two} Y. Cai and S.-Y. Zhu, Phys. Rev. E {\bf 71,} 056607 (2005).

\bibitem{Pittman} T. B. Pittman, Y. H. Shih, D. V. Strekalov, and A. V. Sergienko, Phys. Rev. A {\bf 52,} R3429 (1995).

\bibitem{Valencia} A. Valencia, G. Scarcelli, M. D'Angelo, and Y. Shih, Phys. Rev. Lett. {\bf 94,} 063601 (2005).

\bibitem{Ferri} F. Ferri, D. Magatti, A. Gatti, M. Bache, E. Brambilla, and L. A. Lugiato,  Phys. Rev. Lett. {\bf 94,} 183602 (2005).

\bibitem{Meyers:ToySoldier} R. Meyers, K. S. Deacon, and Y. Shih, Phys. Rev. A {\bf 77,} 041801(R) (2008).

\bibitem{ErkmenShapiro:GI} B. I. Erkmen and J. H. Shapiro, Phys. Rev. A {\bf 77,} 043809 (2008).

\bibitem{footnote1} Earlier theoretical treatments of classical and quantum ghost imaging have also appeared, see, e.g., \cite{Gatti:two,Gatti:three}.

\bibitem{Gatti:four} A. Gatti, M. Bache, D. Magatti, E. Brambilla, F. Ferri and L. A. Lugiato J. Mod. Opt. \textbf{53,} 739 (2006).

\bibitem{Scarcelli} G. Scarcelli, V. Berardi, and Y. Shih, Phys. Rev. Lett. {\bf 96,} 063602 (2006).

\bibitem{ChangHan:GISNR} J. Cheng and S-.S. Han, Chin. Phys. Lett. {\bf 22,} 1676--1679 (2005).

\bibitem{SalehTeich:GINoise} B. E. A. Saleh and M. C. Teich, ``Noise in classical and quantum photon-correlation imaging,'' in A. T. Friberg and R. D\"{a}ndliker, eds., \em Advances in Information Optics and Photonics\/\rm, {\bf PM183} (SPIE Press, Bellingham, 2008), chapter 21.

\bibitem{footnote2} We consider a nonzero distance between the source and object planes keeping in mind remote sensing applications. Furthermore, we have shown in \cite{ErkmenShapiro:GI} that a separation between the object plane and the bucket detector has no impact on resolution (and therefore SNR) if appropriate relay optics are inserted in the signal arm. Therefore, we do not consider any distance between the object and the bucket detector.

\bibitem{YuenShapiroPartI}H. P. Yuen and J. H. Shapiro, IEEE Trans.\ Inform.\ Theory {\bf 24,} 657 (1978).

\bibitem{footnote3} Several other ghost imaging configurations have been reported. In particular, ghost imaging has been performed in reflection \cite{Meyers:ToySoldier}, and it has been performed with lenses in both arms \cite{Ferri}.  Our SNR analysis can be applied to these cases---which we shall not treat---as well those we shall consider explicitly.

\bibitem{footnote4}The {\sc ac}-coupling must resemble a high-quality zero-frequency notch filter, so that the {\sc dc} component is removed with negligible attenuation of the desired baseband photocurrent fluctuations.  The filter $H_B(\Omega)$, including this {\sc ac}-coupling, will be assumed to be within the photodetector blocks shown in Fig.~\ref{GI:propagation}, so that $\langle\hat{\imath}_m(t)\rangle = 0$ for $m=1,2$ for all the field states we shall consider.

\bibitem{footnote5}As in \cite{ErkmenShapiro:GI}, we are using the quantum theory of photodetection 
from H. P. Yuen and J. H. Shapiro, IEEE Trans.\ Inform.\ Theory {\bf 26}, 78 (1980) to represent classical photocurrents by quantum operators whose quantum-measurement statistics coincide with those of the classical photocurrents.

\bibitem{Mandel} L. Mandel and E. Wolf, \em Optical Coherence and Quantum Optics,\/\rm\ (Cambridge Univ., Cambridge, 1995), chapters~4,~10,~11.

\bibitem{Wozencraft} J. M. Wozencraft and I. M. Jacobs, \emph{Principles of Communication Engineering}, (Wiley, New York, 1965), chapter 3.

\bibitem{footnote6} This is the same parametric dependence reported earlier in \cite{Gatti:four}, without derivation.

\bibitem{footnote7} See \cite{ErkmenShapiro:GI} for additional details about coherence propagation into the far field.  Note that (39) in \cite{ErkmenShapiro:GI} has, without comment, suppressed the quadratic phase factors that are irrelevant to formation of the lensless ghost image.  

\bibitem{footnote8} Narrowband classical phase-sensitive light can be generated by dividing a continuous-wave laser beam with a 50-50 beam splitter, and imposing complex-conjugate modulations on the two beams (for example, by using telecom-grade electro-optic modulators). To obtain broadband classical phase-sensitive light one could use spontaneous parametric downconversion with thermal-state signal and idler inputs, such that the joint signal and idler output state is a classical Gaussian state with a phase-sensitive cross correlation.

\bibitem{WongKimShapiro} F. N. C. Wong, T. H. Kim, and J. H. Shapiro, Laser Physics {\bf 16,} 1517 (2006).

\bibitem{Brambilla:SPDC} E. Brambilla, A. Gatti, M. Bache, and L. A. Lugiato, Phys. Rev. A {\bf 69,} 023802 (2004).

\bibitem{footnote9} The SPDC output field operators presented herein are derived from quantized coupled-mode equations using the typical nondepleting plane-wave pump approximation. Furthermore, the transverse boundary effects within the crystal have been ignored, and unimportant global phase factors have been omitted.

\bibitem{footnote10} Although the exact solution of the coupled-mode equations and the boundary conditions at the input facet of the nonlinear crystal does not lead to a Gaussian $\nu(\kvec, \Omega)$, this assumption facilitates an analytic treatment without compromising the fundamental physics we are after.

\bibitem{footnote11}Here we present near-field results, but the corresponding far-field results are easily developed, as discussed in Sec.~II.

\bibitem{footnote12} It is necessary to utilize photon-number resolving detectors in order to reap the advantages ascribed to this high-flux, low-brightness regime.


\end{thebibliography}
\end{document}